\documentclass[letterpaper,compsoc,twoside]{IEEEtran}
\usepackage{fixltx2e} 
\usepackage{cmap} 
\usepackage{ifthen}
\usepackage[T1]{fontenc}
\usepackage[utf8]{inputenc}
\usepackage{amsmath}

\usepackage[font={small,it},labelfont=bf]{caption}
\usepackage{float}

\usepackage{longtable,ltcaption,array}
\newlength{\DUtablewidth} 

\pdfoutput=1
\usepackage{scipy}
\makeatletter
\def\PY@reset{\let\PY@it=\relax \let\PY@bf=\relax%
    \let\PY@ul=\relax \let\PY@tc=\relax%
    \let\PY@bc=\relax \let\PY@ff=\relax}
\def\PY@tok#1{\csname PY@tok@#1\endcsname}
\def\PY@toks#1+{\ifx\relax#1\empty\else%
    \PY@tok{#1}\expandafter\PY@toks\fi}
\def\PY@do#1{\PY@bc{\PY@tc{\PY@ul{%
    \PY@it{\PY@bf{\PY@ff{#1}}}}}}}
\def\PY#1#2{\PY@reset\PY@toks#1+\relax+\PY@do{#2}}

\expandafter\def\csname PY@tok@gd\endcsname{\def\PY@tc##1{\textcolor[rgb]{0.63,0.00,0.00}{##1}}}
\expandafter\def\csname PY@tok@gu\endcsname{\let\PY@bf=\textbf\def\PY@tc##1{\textcolor[rgb]{0.50,0.00,0.50}{##1}}}
\expandafter\def\csname PY@tok@gt\endcsname{\def\PY@tc##1{\textcolor[rgb]{0.00,0.27,0.87}{##1}}}
\expandafter\def\csname PY@tok@gs\endcsname{\let\PY@bf=\textbf}
\expandafter\def\csname PY@tok@gr\endcsname{\def\PY@tc##1{\textcolor[rgb]{1.00,0.00,0.00}{##1}}}
\expandafter\def\csname PY@tok@cm\endcsname{\let\PY@it=\textit\def\PY@tc##1{\textcolor[rgb]{0.25,0.50,0.56}{##1}}}
\expandafter\def\csname PY@tok@vg\endcsname{\def\PY@tc##1{\textcolor[rgb]{0.73,0.38,0.84}{##1}}}
\expandafter\def\csname PY@tok@m\endcsname{\def\PY@tc##1{\textcolor[rgb]{0.13,0.50,0.31}{##1}}}
\expandafter\def\csname PY@tok@mh\endcsname{\def\PY@tc##1{\textcolor[rgb]{0.13,0.50,0.31}{##1}}}
\expandafter\def\csname PY@tok@cs\endcsname{\def\PY@tc##1{\textcolor[rgb]{0.25,0.50,0.56}{##1}}\def\PY@bc##1{\setlength{\fboxsep}{0pt}\colorbox[rgb]{1.00,0.94,0.94}{\strut ##1}}}
\expandafter\def\csname PY@tok@ge\endcsname{\let\PY@it=\textit}
\expandafter\def\csname PY@tok@vc\endcsname{\def\PY@tc##1{\textcolor[rgb]{0.73,0.38,0.84}{##1}}}
\expandafter\def\csname PY@tok@il\endcsname{\def\PY@tc##1{\textcolor[rgb]{0.13,0.50,0.31}{##1}}}
\expandafter\def\csname PY@tok@go\endcsname{\def\PY@tc##1{\textcolor[rgb]{0.20,0.20,0.20}{##1}}}
\expandafter\def\csname PY@tok@cp\endcsname{\def\PY@tc##1{\textcolor[rgb]{0.00,0.44,0.13}{##1}}}
\expandafter\def\csname PY@tok@gi\endcsname{\def\PY@tc##1{\textcolor[rgb]{0.00,0.63,0.00}{##1}}}
\expandafter\def\csname PY@tok@gh\endcsname{\let\PY@bf=\textbf\def\PY@tc##1{\textcolor[rgb]{0.00,0.00,0.50}{##1}}}
\expandafter\def\csname PY@tok@ni\endcsname{\let\PY@bf=\textbf\def\PY@tc##1{\textcolor[rgb]{0.84,0.33,0.22}{##1}}}
\expandafter\def\csname PY@tok@nl\endcsname{\let\PY@bf=\textbf\def\PY@tc##1{\textcolor[rgb]{0.00,0.13,0.44}{##1}}}
\expandafter\def\csname PY@tok@nn\endcsname{\let\PY@bf=\textbf\def\PY@tc##1{\textcolor[rgb]{0.05,0.52,0.71}{##1}}}
\expandafter\def\csname PY@tok@no\endcsname{\def\PY@tc##1{\textcolor[rgb]{0.38,0.68,0.84}{##1}}}
\expandafter\def\csname PY@tok@na\endcsname{\def\PY@tc##1{\textcolor[rgb]{0.25,0.44,0.63}{##1}}}
\expandafter\def\csname PY@tok@nb\endcsname{\def\PY@tc##1{\textcolor[rgb]{0.00,0.44,0.13}{##1}}}
\expandafter\def\csname PY@tok@nc\endcsname{\let\PY@bf=\textbf\def\PY@tc##1{\textcolor[rgb]{0.05,0.52,0.71}{##1}}}
\expandafter\def\csname PY@tok@nd\endcsname{\let\PY@bf=\textbf\def\PY@tc##1{\textcolor[rgb]{0.33,0.33,0.33}{##1}}}
\expandafter\def\csname PY@tok@ne\endcsname{\def\PY@tc##1{\textcolor[rgb]{0.00,0.44,0.13}{##1}}}
\expandafter\def\csname PY@tok@nf\endcsname{\def\PY@tc##1{\textcolor[rgb]{0.02,0.16,0.49}{##1}}}
\expandafter\def\csname PY@tok@si\endcsname{\let\PY@it=\textit\def\PY@tc##1{\textcolor[rgb]{0.44,0.63,0.82}{##1}}}
\expandafter\def\csname PY@tok@s2\endcsname{\def\PY@tc##1{\textcolor[rgb]{0.25,0.44,0.63}{##1}}}
\expandafter\def\csname PY@tok@vi\endcsname{\def\PY@tc##1{\textcolor[rgb]{0.73,0.38,0.84}{##1}}}
\expandafter\def\csname PY@tok@nt\endcsname{\let\PY@bf=\textbf\def\PY@tc##1{\textcolor[rgb]{0.02,0.16,0.45}{##1}}}
\expandafter\def\csname PY@tok@nv\endcsname{\def\PY@tc##1{\textcolor[rgb]{0.73,0.38,0.84}{##1}}}
\expandafter\def\csname PY@tok@s1\endcsname{\def\PY@tc##1{\textcolor[rgb]{0.25,0.44,0.63}{##1}}}
\expandafter\def\csname PY@tok@gp\endcsname{\let\PY@bf=\textbf\def\PY@tc##1{\textcolor[rgb]{0.78,0.36,0.04}{##1}}}
\expandafter\def\csname PY@tok@sh\endcsname{\def\PY@tc##1{\textcolor[rgb]{0.25,0.44,0.63}{##1}}}
\expandafter\def\csname PY@tok@ow\endcsname{\let\PY@bf=\textbf\def\PY@tc##1{\textcolor[rgb]{0.00,0.44,0.13}{##1}}}
\expandafter\def\csname PY@tok@sx\endcsname{\def\PY@tc##1{\textcolor[rgb]{0.78,0.36,0.04}{##1}}}
\expandafter\def\csname PY@tok@bp\endcsname{\def\PY@tc##1{\textcolor[rgb]{0.00,0.44,0.13}{##1}}}
\expandafter\def\csname PY@tok@c1\endcsname{\let\PY@it=\textit\def\PY@tc##1{\textcolor[rgb]{0.25,0.50,0.56}{##1}}}
\expandafter\def\csname PY@tok@kc\endcsname{\let\PY@bf=\textbf\def\PY@tc##1{\textcolor[rgb]{0.00,0.44,0.13}{##1}}}
\expandafter\def\csname PY@tok@c\endcsname{\let\PY@it=\textit\def\PY@tc##1{\textcolor[rgb]{0.25,0.50,0.56}{##1}}}
\expandafter\def\csname PY@tok@mf\endcsname{\def\PY@tc##1{\textcolor[rgb]{0.13,0.50,0.31}{##1}}}
\expandafter\def\csname PY@tok@err\endcsname{\def\PY@bc##1{\setlength{\fboxsep}{0pt}\fcolorbox[rgb]{1.00,0.00,0.00}{1,1,1}{\strut ##1}}}
\expandafter\def\csname PY@tok@kd\endcsname{\let\PY@bf=\textbf\def\PY@tc##1{\textcolor[rgb]{0.00,0.44,0.13}{##1}}}
\expandafter\def\csname PY@tok@ss\endcsname{\def\PY@tc##1{\textcolor[rgb]{0.32,0.47,0.09}{##1}}}
\expandafter\def\csname PY@tok@sr\endcsname{\def\PY@tc##1{\textcolor[rgb]{0.14,0.33,0.53}{##1}}}
\expandafter\def\csname PY@tok@mo\endcsname{\def\PY@tc##1{\textcolor[rgb]{0.13,0.50,0.31}{##1}}}
\expandafter\def\csname PY@tok@mi\endcsname{\def\PY@tc##1{\textcolor[rgb]{0.13,0.50,0.31}{##1}}}
\expandafter\def\csname PY@tok@kn\endcsname{\let\PY@bf=\textbf\def\PY@tc##1{\textcolor[rgb]{0.00,0.44,0.13}{##1}}}
\expandafter\def\csname PY@tok@o\endcsname{\def\PY@tc##1{\textcolor[rgb]{0.40,0.40,0.40}{##1}}}
\expandafter\def\csname PY@tok@kr\endcsname{\let\PY@bf=\textbf\def\PY@tc##1{\textcolor[rgb]{0.00,0.44,0.13}{##1}}}
\expandafter\def\csname PY@tok@s\endcsname{\def\PY@tc##1{\textcolor[rgb]{0.25,0.44,0.63}{##1}}}
\expandafter\def\csname PY@tok@kp\endcsname{\def\PY@tc##1{\textcolor[rgb]{0.00,0.44,0.13}{##1}}}
\expandafter\def\csname PY@tok@w\endcsname{\def\PY@tc##1{\textcolor[rgb]{0.73,0.73,0.73}{##1}}}
\expandafter\def\csname PY@tok@kt\endcsname{\def\PY@tc##1{\textcolor[rgb]{0.56,0.13,0.00}{##1}}}
\expandafter\def\csname PY@tok@sc\endcsname{\def\PY@tc##1{\textcolor[rgb]{0.25,0.44,0.63}{##1}}}
\expandafter\def\csname PY@tok@sb\endcsname{\def\PY@tc##1{\textcolor[rgb]{0.25,0.44,0.63}{##1}}}
\expandafter\def\csname PY@tok@k\endcsname{\let\PY@bf=\textbf\def\PY@tc##1{\textcolor[rgb]{0.00,0.44,0.13}{##1}}}
\expandafter\def\csname PY@tok@se\endcsname{\let\PY@bf=\textbf\def\PY@tc##1{\textcolor[rgb]{0.25,0.44,0.63}{##1}}}
\expandafter\def\csname PY@tok@sd\endcsname{\let\PY@it=\textit\def\PY@tc##1{\textcolor[rgb]{0.25,0.44,0.63}{##1}}}

\def\PYZbs{\char`\\}
\def\PYZus{\char`\_}

\def\PYZsh{\char`\#}

\def\PYZhy{\char`\-}

\def\PYZdq{\char`\"}

\makeatother

\providecommand*{\DUprovidelength}[2]{
  \ifthenelse{\isundefined{#1}}{\newlength{#1}\setlength{#1}{#2}}{}
}

\providecommand*{\DUrole}[2]{%
  \ifcsname DUrole#1\endcsname%
    \csname DUrole#1\endcsname{#2}%
  \else
    \ifcsname docutilsrole#1\endcsname%
      \csname docutilsrole#1\endcsname{#2}%
    \else%
      #2%
    \fi%
  \fi%
}

\DUprovidelength{\DUlineblockindent}{2.5em}
\ifthenelse{\isundefined{\DUlineblock}}{
  \newenvironment{DUlineblock}[1]{%
    \list{}{\setlength{\partopsep}{\parskip}
            \addtolength{\partopsep}{\baselineskip}
            \setlength{\topsep}{0pt}
            \setlength{\itemsep}{0.15\baselineskip}
            \setlength{\parsep}{0pt}
            \setlength{\leftmargin}{#1}}
    \raggedright
  }
  {\endlist}
}{}

\ifthenelse{\isundefined{\hypersetup}}{
  \usepackage[colorlinks=true,linkcolor=blue,urlcolor=blue]{hyperref}
  \urlstyle{same} 
}{}

\begin{document}
\newcounter{footnotecounter}\title{pyFRET: A Python Library for Single Molecule Fluorescence Data Analysis}\author{Rebecca R. Murphy$^{\setcounter{footnotecounter}{1}\fnsymbol{footnotecounter}\setcounter{footnotecounter}{2}\fnsymbol{footnotecounter}}$%
          \setcounter{footnotecounter}{1}\thanks{\fnsymbol{footnotecounter} %
          Corresponding author: \protect\href{mailto:rrm33@cam.ac.uk}{rrm33@cam.ac.uk}}\setcounter{footnotecounter}{2}\thanks{\fnsymbol{footnotecounter} Department of Chemistry, University of Cambridge}, Sophie E. Jackson$^{\setcounter{footnotecounter}{2}\fnsymbol{footnotecounter}}$, David Klenerman$^{\setcounter{footnotecounter}{2}\fnsymbol{footnotecounter}}$\thanks{%

          \noindent%
          Copyright\,\copyright\,2014 Rebecca R. Murphy et al. This is an open-access article distributed under the terms of the Creative Commons Attribution License, which permits unrestricted use, distribution, and reproduction in any medium, provided the original author and source are credited. http://creativecommons.org/licenses/by/3.0/%
        }}\maketitle
          \renewcommand{\leftmark}{PROC. OF THE 7th EUR. CONF. ON PYTHON IN SCIENCE (EUROSCIPY 2014)}
          \renewcommand{\rightmark}{PYFRET: A PYTHON LIBRARY FOR SINGLE MOLECULE FLUORESCENCE DATA ANALYSIS}

\setcounter{page}{59}
\newcommand*{\docutilsroleref}{\ref}
\newcommand*{\docutilsrolelabel}{\label}
\AtEndDocument{\cleardoublepage}
\begin{abstract}Single molecule Förster resonance energy transfer (smFRET) is a powerful experimental technique for studying the properties of individual biological molecules in solution.
However, as adoption of smFRET techniques becomes more widespread, the lack of available software, whether open source or commercial, for data analysis, is becoming a significant issue.
Here, we present pyFRET, an open source Python package for the analysis of data from single-molecule fluorescence experiments from freely diffusing biomolecules.
The package provides methods for the complete analysis of a smFRET dataset, from burst selection and denoising, through data visualisation and model fitting. We provide support for both continuous excitation and alternating laser excitation (ALEX) data analysis.
pyFRET is available as a package downloadable from the Python Package Index (PyPI) under the open source three-clause BSD licence, together with links to extensive documentation and tutorials, including example usage and test data.
Additional documentation including tutorials is hosted independently on ReadTheDocs.
The code is available from the free hosting site Bitbucket.
Through distribution of this software, we hope to lower the barrier for the adoption of smFRET experiments by other research groups and we encourage others to contribute modules for specific analysis needs.\end{abstract}\begin{IEEEkeywords}smFRET, single-molecule, confocal, python, fluorescence\end{IEEEkeywords}

\section{Introduction%
  \label{introduction}%
}

Förster resonance energy transfer (FRET) \cite{Forster48} is a physical process that allows the study of molecular interactions and intramolecular distances.
FRET is the non-radiative transfer of energy between two fluorescent molecules, where the fraction of energy transferred varies with the sixth power of the inter-fluorophore distance, providing an extremely sensitive readout of the distance between two fluorophores.
Since its first demonstration, \cite{ha96}, single-molecule FRET (smFRET) has grown in popularity as a tool to investigate the structure and dynamics of biomolecules diffusing in solution \cite{haran03}, \cite{schuler02}, \cite{weiss00}.

In a smFRET experiment, biological molecules are labelled with two fluorescent dyes, selected such that the emission spectrum of one dye (the donor, D) overlaps with the excitation spectrum of the other (the acceptor, A).
When the donor and acceptor are physically close in space, exciting the donor dye can result in emission from the acceptor dye, where the proportion of emission from the acceptor and donor, known as the FRET Efficiency, E, depends on the distance, r between the two dyes and $R_0$:, the Förster distance, a dye dependent constant that describes the dye separation at which 50\% energy transfer is achieved.\begin{equation*}
E = \frac{1}{1 + (\frac{r}{R_0})^6}
\end{equation*}Experimentally, a laser beam is focused through a highly dilute solution of labelled molecules onto a diffraction-limited focal point. When a labelled molecule diffuses through the laser beam, the donor dye is excited and photons are emitted (Fig. \DUrole{ref}{microscope}).
Emitted photons are collected through the objective and separated by a dichroic mirror into donor and acceptor photons.
The ratio of acceptor to donor photons allows calculation of E for that fluorescent event:\begin{equation*}
E = \frac{n_A}{n_A + \gamma \cdot n_D}
\end{equation*}for n\textsubscript{A} and n\textsubscript{D} photons in the acceptor and donor channels respectively and $\gamma$ an experimentally determined instrument-dependent factor that corrects for unequal detection efficiencies.
The thousands of fluorescent events collected during an experiment are used to construct a FRET efficiency histogram, which is typically fitted with a gaussian distribution to identify populations of fluorescent species \cite{ha96}, \cite{nir11}.\begin{figure}[]\noindent\makebox[\columnwidth][c]{\includegraphics[width=\columnwidth]{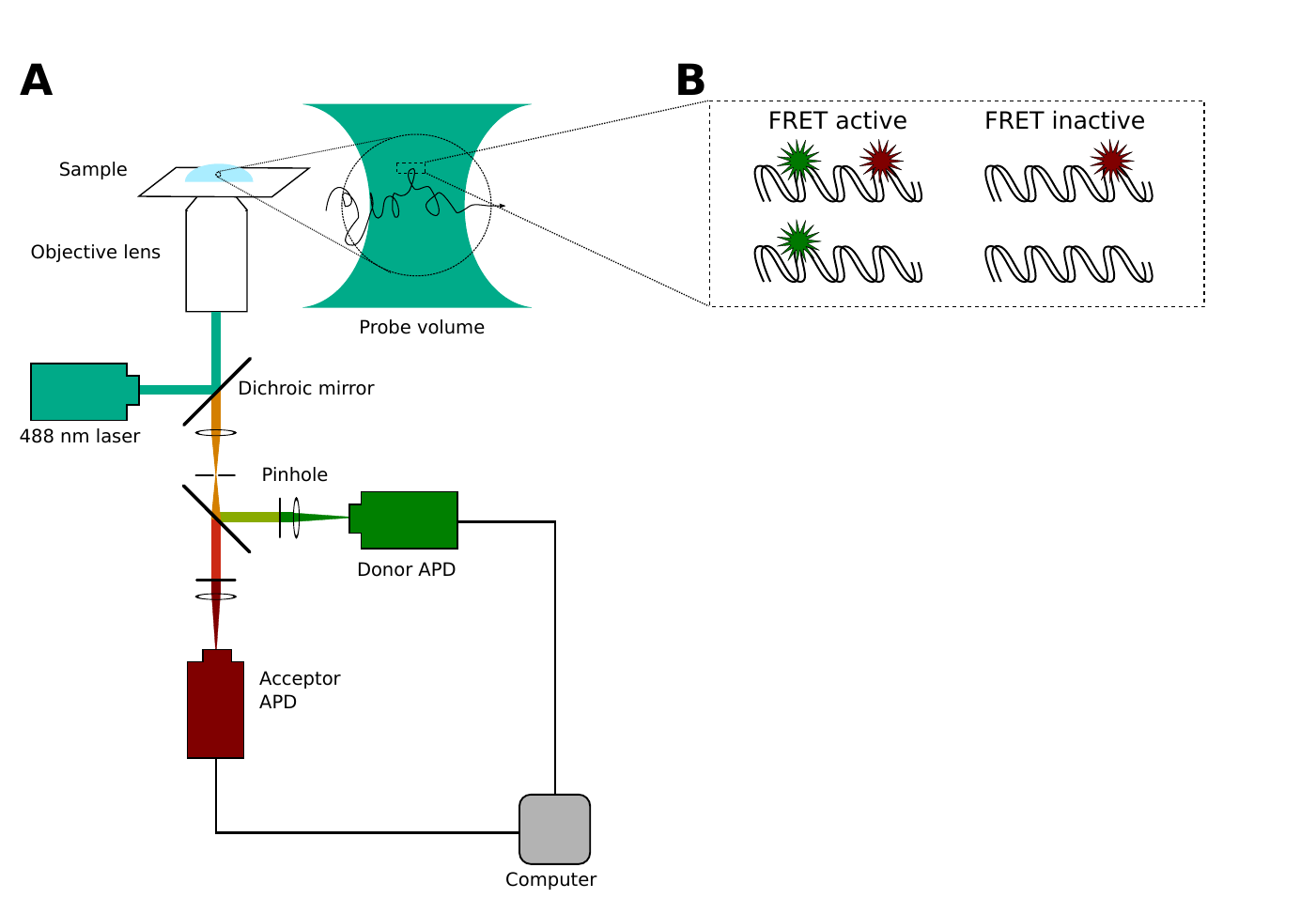}}
\caption{Instrumentation for a smFRET experiment. \DUrole{label}{microscope}}
\end{figure}

A smFRET experiment involves several computational challenges.
Bursts of fluorescence emission, corresponding to a molecule diffusing through the laser beam, must be identified against a noisy background and identified bursts must be denoised.
Correction of photobleaching effects, donor-acceptor crosstalk and other photophysical artifacts must also be applied to get accurate intramolecular distance information. Multiple methods of burst selection and analysis have been developed and applied to the analysis of smFRET data \cite{weiss00}, \cite{deniz01}, \cite{gell06}, \cite{nir06}, \cite{kapanidis05}, \cite{muller05}, \cite{doose07}, \cite{kudryavtsev2012}, \cite{eggeling01}.
However, software for analysis of smFRET data has thus far been developed on an ad hoc basis, with individual groups preparing and maintaining their own analysis scripts.

This method of software development has created several problems for the smFRET research community that are typical of research programming projects \cite{wilson06}, \cite{merali10}.

Firstly, there is the problem of \textquotedbl{}reinventing the wheel\textquotedbl{} \cite{mirams13}. Within smFRET research groups, programming ability is not a standard skill, despite the need for sophisticated data analysis and use of custom data collection hardware.
It is common for researchers with programming skills to maintain their own series of data-analysis scripts which may be wholly dependent on particular hardware tools or analysis packages.
Other researchers, who may lack the skills to maintain and develop even simple scripts, are dependent on black-box techniques provided by their colleagues. Consequently, data analysis is dependent on scripts written and maintained by just a few researchers.
Loss of programming expertise when these team members leave can result in significant difficulties for the remaining group members, who are then dependent on poorly documented code that they do not fully understand how to use.
Furthermore, the lack of available open source software often requires new researchers in the field of smFRET to completely reimplement standard analysis techniques in order to become independently productive.

Secondly, the need for many researchers to develop and maintain their own analyis tools has significant impact on research productivity.
The requirement to reimplement standard analysis techniques consumes valuable time that could better be used in experimental research or in developing and benchmarking improved analysis tools.
Furthermore, most researchers have no formal training in software engineering, with the result that analysis software can vary hugely in quality and is frequently poorly documented and maintained, making it difficult for other researchers to understand and use.
New analysis scipts are often added in an ad hoc manner, with the result that straighforward tasks are performed using an unwieldy mess of spaghetti code, transforming simple modifications into complex undertakings requiring significant time investment.
Poorly maintained code adds an additional barrier to open sharing of resources as groups are embarrassed to share low-quality software.

Finally, there is the issue of research reproducibility. Different research groups use widely differing tools to complete similar tasks.
New methods of data collection and analysis are frequently developed \cite{kapanidis05}, \cite{nir06}, \cite{sisamakis2010}.
However, when software is not released to the community, it is difficult for researchers, who must often implement poorly described methodologies entirely from scratch, to verify results or to adopt new techniques in their own research. As a consequence, new techniques are poorly benchmarked, making it difficult to understand whether a new analysis adds quality or merely complexity, whilst adoption of useful new methods is relatively slow.
These three issues of productivity, reliability and reproducibility, all linked to the problem of poorly maintained softwared and lack of software development skills, are now becoming a key bottleneck in smFRET research.

We have developed pyFRET, a fully open source library, written in Python, for the analysis of smFRET data.
To our knowledge, this is the first open source code ever released by the smFRET research community.
Our library aims to address the issues described above by providing a simple toolkit for smFRET data analysis.
pyFRET is a small library, consisting of just 700 lines of Python code (including inline comments).
However, it contains functions for all key steps in analysis of smFRET data, including burst selection; cross-talk subtraction and burst denoising; data visualisation; and construction and simple fitting of FRET efficiency histograms.
In providing this toolkit to the smFRET research community, we hope to facilitate the wider adoption of smFRET techniques in biological research as well as to provide a framework for open communication about and sharing of data analsyis tools.

\section{Design and Implementation%
  \label{design-and-implementation}%
}

\subsection{Implementation%
  \label{implementation}%
}

pyFRET provides two key classes for manipulation of smFRET data.
The FRET data object describes two fluorescence channels, corresponding to time-bins containing photons collected from donor (the donor channel, D) and acceptor (the acceptor channel, A) fluorophores.
The ALEX data object describes four fluorescence channels, corresponding to the four temporal states in a smFRET experiment using Alternating Laser Excitation (ALEX), namely the donor channel when the donor laser is switched on (D\_D); the donor channel when the acceptor laser is switched on (D\_A); the acceptor channel when the donor laser is on (A\_D); and the acceptor channel when the acceptor laser is on (A\_A).
These data channels are implemented as numpy arrays, allowing efficient computation and selection operations.

The data analysis workflow is illustrated in Figure \DUrole{ref}{fig1workflow}.
Following initialization of data objects, background subtraction, event selection, cross-talk correction and calculation of the FRET efficiency can each be performed with a single call to a pyFRET function.

pyFRET provides built-in functions to generate the most common plot types used in the field.
For example, proximity ratio histograms, which allow identification and analysis of different fluorescent populations, can be generated using the proximity\_ratio method (Fig. \DUrole{ref}{fig2plots} A).
For ALEX data, scatter plots with projected histograms can be constructed (Fig. \DUrole{ref}{fig2plots} B).
Further plotting options are shown in Fig. \DUrole{ref}{fig2plots} C and D.\begin{figure*}[]\noindent\makebox[\textwidth][c]{\includegraphics[width=\columnwidth]{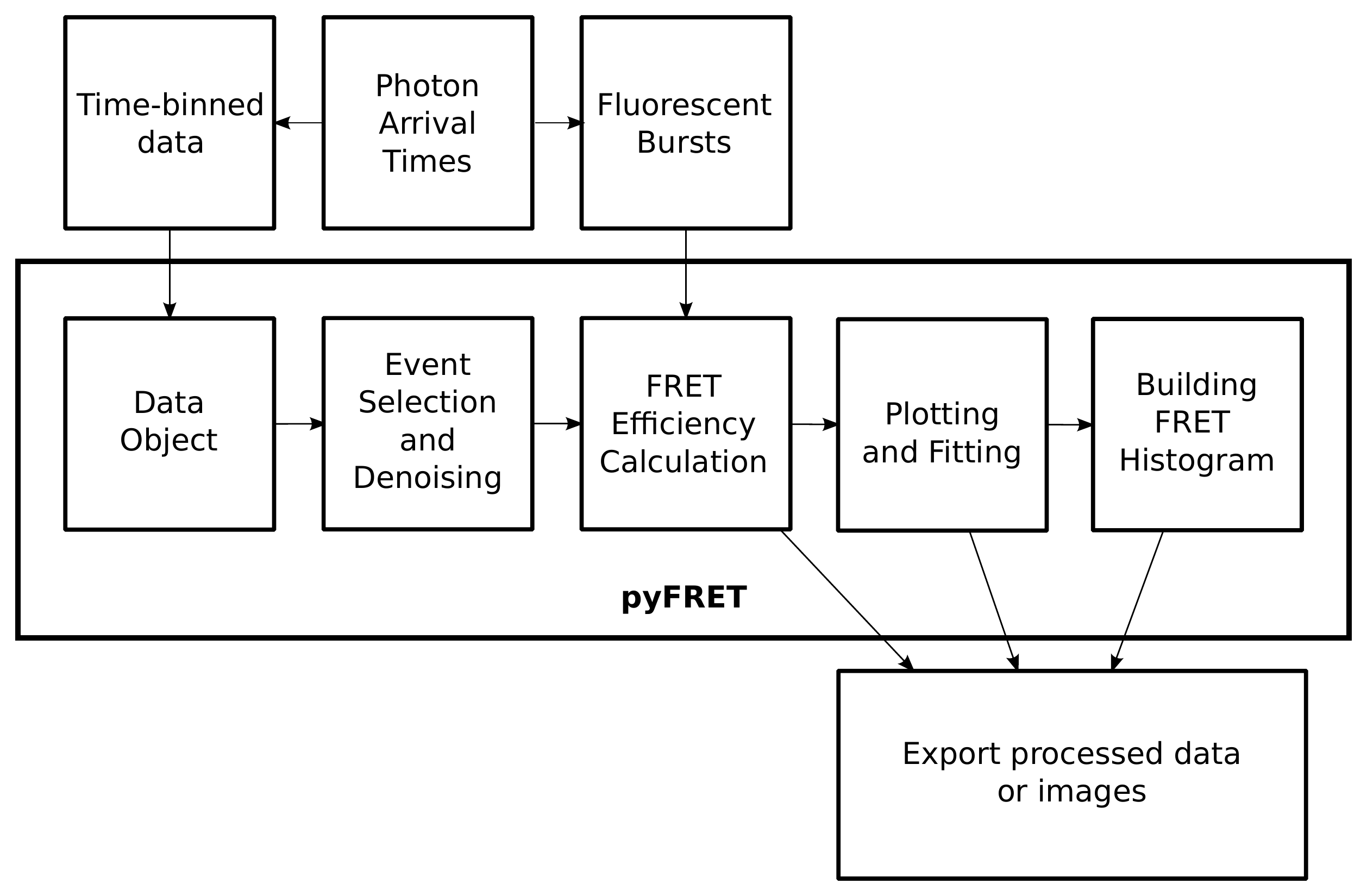}}
\caption{Typical workflow for data analysis using pyFRET. \DUrole{label}{fig1workflow}}
\end{figure*}\begin{figure}[]\noindent\makebox[\columnwidth][c]{\includegraphics[width=\columnwidth]{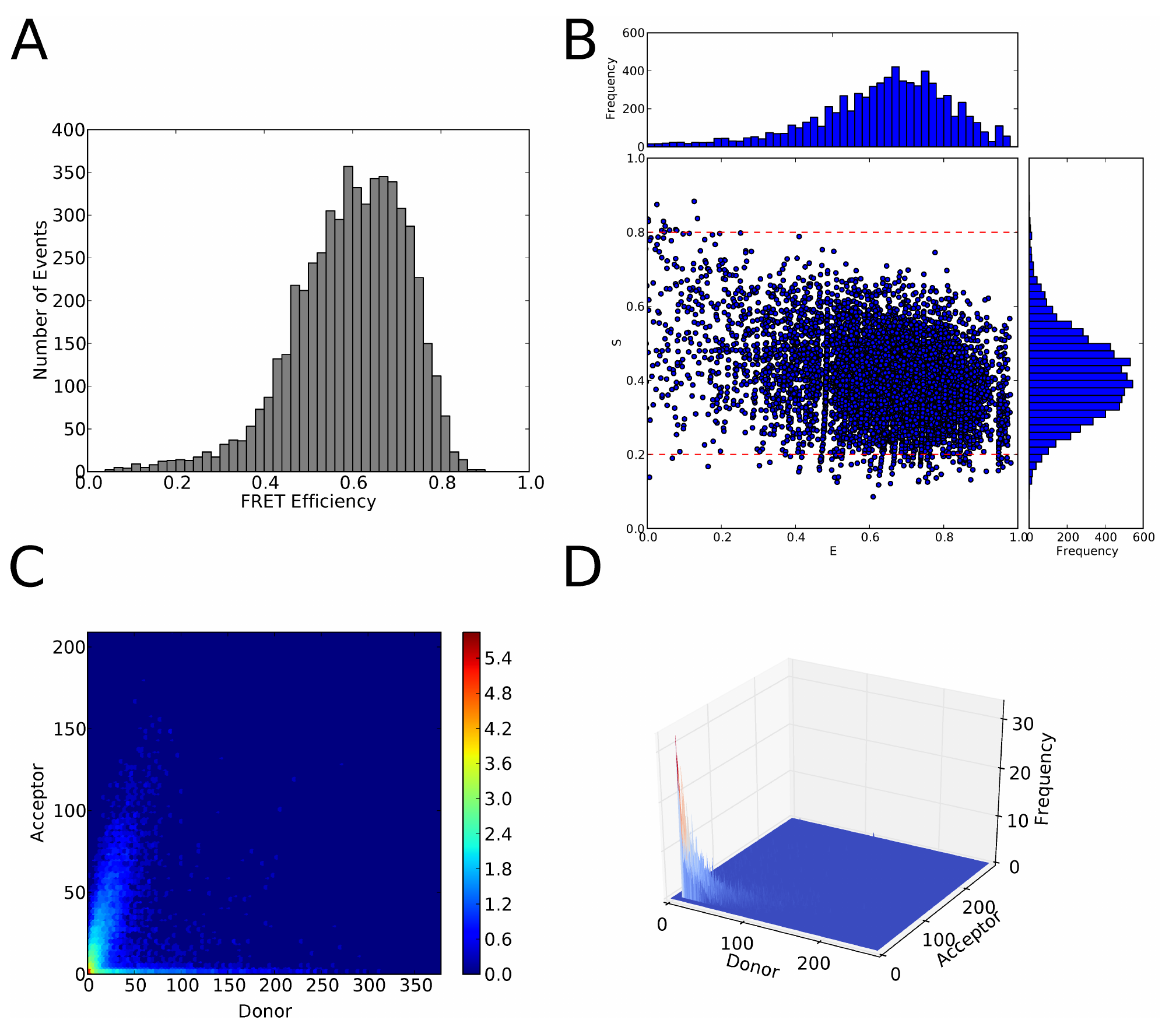}}
\caption{Figures made using pyFRET. A) A Proximity Ratio histogram. B) A scatter-plot of FRET efficiency and fluorophore stoichiometry from ALEX data. C) A heatmap of event frequencies.  D) A 3D plot of event frequencies. \DUrole{label}{fig2plots}}
\end{figure}

\subsection{Compatibility Considerations%
  \label{compatibility-considerations}%
}

pyFRET is written in Python. Both Python 2 (v2.7) and Python 3 (v3.3) are supported.
pyFRET requires three further Python libraries,  namely numpy \cite{numpy} and scipy \cite{scipy} for data manipulation, and matplotlib \cite{matplotlib} for data visualisation.
Installation of pyFRET using the pip install method supported by PyPI will facilitate automatic installation of these packages if they are not already included in your Python build.

The lack of open source software in the smFRET community has led to a proliferation of esoteric file-types used for data collection and storage.
To make pyFRET as usable as possible for a wide range of smFRET researchers, the pyFRET data structures can be initialised using arrays of time-binned photons. The tutorial also demonstrates using pyFRET's file-parsing functions scripts to create pyFRET objects from common filetypes.

pyFRET currently provides basic tools for analysis and visualisation of smFRET data.
In the interest of providing the pyFRET infrastructure to smFRET researchers at an early stage, we are choosing to release our software at a relatively early stage of development.
pyFRET provides a complete tool-chain for analysis of time-binned smFRET data, but does not currently include a burst-search algorithm for identification of fluorescent bursts from photon arrival times \cite{nir06}. Fluorescent bursts identified using a burst search algorithm can be analysed using pyFRET by initialising a pyFRET data object from the paired burst photon frequencies.   Denoising and cross-talk correction is achieved in exactly the same manner as for time-binned data, but thresholding is not required.
We encourage researchers who wish to use pyFRET in its current implementation for data visualisation and analysis, but whose data consists of time-stamped photon arrivals to apply their own burst selection algorithms to generate arrays of fluorescent bursts that can be manipulated using pyFRET methods. We also welcome pull requests and contributions to the bitbucket repository.

\section{Experimental Methods%
  \label{experimental-methods}%
}

We tested the pyFRET library using DNA duplexes dual-labelled with Alexa Fluor 488 and Alexa Fluor 647. The duplex sequences and labelling sites are shown in Tables \DUrole{ref}{tab-donor} (Donor strand) and \DUrole{ref}{tab-acceptor} (Acceptor strands).
DNA duplexes were prepared by mixing a 1.1 molar excess of the appropriate acceptor strand with the donor strand, heating to 95 C for 10 minutes, then gradual cooling to room temperature.
FRET data were collected for 15 minutes using continuous excitation at 488 nm and binned in intervals of 1 ms.
ALEX data were collected for 15 minutes using alternating excitation at 488 and 640 nm, with a modulation rate of 0.1 ms, a dead-time of 0.1 $\mu$ s and a delay compensation of 3 $\mu$ s.
ALEX data were then binned in intervals of 1 ms. The scripts and configuration files used to analyse these data using pyFRET can be found in the \textquotedbl{}bin\textquotedbl{} folder of the pyFRET repository.\begin{table*}
\setlength{\DUtablewidth}{0.8\linewidth}
\begin{longtable*}[c]{|p{0.207\DUtablewidth}|p{0.733\DUtablewidth}|}
\hline

Donor Construct & 

Sequence \\
\hline

Donor & 

TACTGCCTTTCTGTATCGC5TATCGCGTAGTTACCTGCCTTGCATAGCCACTCATAGCCT \\
\hline
\end{longtable*}
\caption{DNA sequence of the donor-labelled strand, where 5 is a deoxy-T nucleotide, labelled with Alexa Fluor 488 at the C6 amino position  \DUrole{label}{tab-donor}}\end{table*}\begin{table*}
\setlength{\DUtablewidth}{0.8\linewidth}
\begin{longtable*}[c]{|p{0.156\DUtablewidth}|p{0.703\DUtablewidth}|}
\hline

Separation & 

Acceptor Sequence \\
\hline

4 & 

AGGCTATGAGTGGCTATGCAAGGCAGGTAACTACGCGATAAGCGA6 \\
\hline

6 & 

AGGCTATGAGTGGCTATGCAAGGCAGGTAACTACGCGATAAGCGATA6 \\
\hline

8 & 

AGGCTATGAGTGGCTATGCAAGGCAGGTAACTACGCGATAAGCGATACA6 \\
\hline

10 & 

AGGCTATGAGTGGCTATGCAAGGCAGGTAACTACGCGATAAGCGATACAGA6 \\
\hline

12 & 

AGGCTATGAGTGGCTATGCAAGGCAGGTAACTACGCGATAAGCGATACAGAAA6 \\
\hline
\end{longtable*}
\caption{Preparing the dual-labelled dsDNA. An acceptor-labelled ssDNA, with the sequence shown was annealed to the indicated donor construct, to yield a dual-labelled construct with the labels separated by the given number of base pairs. In the displayed acceptor-strand sequences, 6 is a deoxy-T nucleotide, labelled with Alexa Fluor 647 at the C6 amino position.. \DUrole{label}{tab-acceptor}}\end{table*}

\section{Results%
  \label{results}%
}

As an example of the analysis that can be performed using pyFRET, we collected data from dual-labelled DNA duplexes with various dye-dye separation distances, using both FRET and ALEX excitation patterns.
We then analysed the data using the pyFRET analysis pipeline.
Timebins were background corrected and events were selected using a fixed threshold.
FRET efficiency histograms were constructed and fitted to a single gaussian distribution.
The mean FRET efficiencies were then plotted against the dye separation distance to show the characteristic sigmoidal curve.
Results of the analysis are show in Fig. \DUrole{ref}{fig3FRET} (FRET) and Fig. \DUrole{ref}{fig3ALEX} (ALEX).
An example analysis script to produce a fitted smFRET histogram is shown below. Here, the parameters auto\_donor, auto\_acceptor, cross\_DtoA, cross\_AtoD and g\_factor are user-supplied experimentally determined correction factors; T\_donor and T\_acceptor are user-supplied thresholds for event selection. Auto\_donor and auto\_acceptor are the background autofluorescence in the donor and acceptor channels respectively; cross\_DtoA and cross\_AtoD are the crosstalk in the acceptor and donor channels, caused by direct excitation of the acceptor dye by the donor laser and the donor dye by the acceptor laser respectively; g\_factor is the correction factor $\gamma$ described above; and T\_donor and T\_acceptor are photon count thresholds above which a time-bin is classified as containing a fluorescent event. Realistic parameter values are shown in the snippet below.
\begin{DUlineblock}{0em}
\item[] 
\end{DUlineblock}
\begin{Verbatim}[commandchars=\\\{\},fontsize=\footnotesize]
\PY{k+kn}{from} \PY{n+nn}{pyFRET} \PY{k+kn}{import} \PY{n}{pyFRET} \PY{k}{as} \PY{n}{pft}

\PY{c}{\PYZsh{} read data}
\PY{n}{my\PYZus{}directory} \PY{o}{=} \PY{l+s}{\PYZdq{}}\PY{l+s}{path/to/my/files}\PY{l+s}{\PYZdq{}}
\PY{n}{list\PYZus{}of\PYZus{}files} \PY{o}{=} \PY{p}{[}\PY{l+s}{\PYZdq{}}\PY{l+s}{file1.csv}\PY{l+s}{\PYZdq{}}\PY{p}{,} \PY{l+s}{\PYZdq{}}\PY{l+s}{file2.csv}\PY{l+s}{\PYZdq{}}\PY{p}{,} \PY{l+s}{\PYZdq{}}\PY{l+s}{file3.csv}\PY{l+s}{\PYZdq{}}\PY{p}{]}
\PY{n}{my\PYZus{}data} \PY{o}{=} \PY{n}{pft}\PY{o}{.}\PY{n}{parse\PYZus{}csv}\PY{p}{(}\PY{n}{my\PYZus{}directory}\PY{p}{,} \PY{n}{list\PYZus{}of\PYZus{}files}\PY{p}{)}

\PY{c}{\PYZsh{} define constants}
\PY{n}{auto\PYZus{}donor} \PY{o}{=} \PY{l+m+mf}{0.3}      \PY{c}{\PYZsh{} background autofluorescence}
\PY{n}{auto\PYZus{}acceptor} \PY{o}{=} \PY{l+m+mf}{0.2}
\PY{n}{T\PYZus{}donor} \PY{o}{=} \PY{l+m+mi}{15}          \PY{c}{\PYZsh{} photon count thresholds}
\PY{n}{T\PYZus{}acceptor} \PY{o}{=} \PY{l+m+mi}{15}
\PY{n}{cross\PYZus{}DtoA} \PY{o}{=} \PY{l+m+mf}{0.05}     \PY{c}{\PYZsh{} cross\PYZhy{}talk}
\PY{n}{cross\PYZus{}AtoD} \PY{o}{=} \PY{l+m+mf}{0.01}
\PY{n}{g\PYZus{}factor} \PY{o}{=} \PY{l+m+mf}{1.0}        \PY{c}{\PYZsh{} detection correction factor}


\PY{c}{\PYZsh{} background correction and event selection}
\PY{n}{my\PYZus{}data}\PY{o}{.}\PY{n}{subtract\PYZus{}bckd}\PY{p}{(}\PY{n}{auto\PYZus{}donor}\PY{p}{,} \PY{n}{auto\PYZus{}acceptor}\PY{p}{)}
\PY{n}{my\PYZus{}data}\PY{o}{.}\PY{n}{threshold\PYZus{}AND}\PY{p}{(}\PY{n}{T\PYZus{}donor}\PY{p}{,} \PY{n}{T\PYZus{}acceptor}\PY{p}{)}
\PY{n}{my\PYZus{}data}\PY{o}{.}\PY{n}{subtract\PYZus{}crosstalk}\PY{p}{(}\PY{n}{cross\PYZus{}DtoA}\PY{p}{,} \PY{n}{cross\PYZus{}AtoD}\PY{p}{)}

\PY{c}{\PYZsh{} make histogram of FRET efficiency and fit}
\PY{n}{my\PYZus{}data}\PY{o}{.}\PY{n}{build\PYZus{}histogram}\PY{p}{(}\PY{n}{filepath}\PY{p}{,} \PY{n}{csvname}\PY{p}{,} \PYZbs{}
  \PY{n}{gamma}\PY{o}{=}\PY{n}{g\PYZus{}factor}\PY{p}{,} \PY{n}{bin\PYZus{}min}\PY{o}{=}\PY{l+m+mf}{0.0}\PY{p}{,} \PY{n}{bin\PYZus{}max}\PY{o}{=}\PY{l+m+mf}{1.0}\PY{p}{,} \PYZbs{}
  \PY{n}{bin\PYZus{}width}\PY{o}{=}\PY{l+m+mf}{0.02}\PY{p}{,} \PY{n}{image}\PY{o}{=}\PY{n+nb+bp}{True}\PY{p}{,} \PY{n}{imgname}\PY{o}{=}\PY{l+s}{\PYZdq{}}\PY{l+s}{my\PYZus{}histogram}\PY{l+s}{\PYZdq{}}\PY{p}{,} \PYZbs{}
  \PY{n}{imgtype}\PY{o}{=}\PY{l+s}{\PYZdq{}}\PY{l+s}{png}\PY{l+s}{\PYZdq{}}\PY{p}{,} \PY{n}{gauss}\PY{o}{=}\PY{n+nb+bp}{True}\PY{p}{,} \PY{n}{gaussname}\PY{o}{=}\PY{l+s}{\PYZdq{}}\PY{l+s}{gaussfit}\PY{l+s}{\PYZdq{}}\PY{p}{)}
\end{Verbatim}
\begin{figure}[]\noindent\makebox[\columnwidth][c]{\includegraphics[width=\columnwidth]{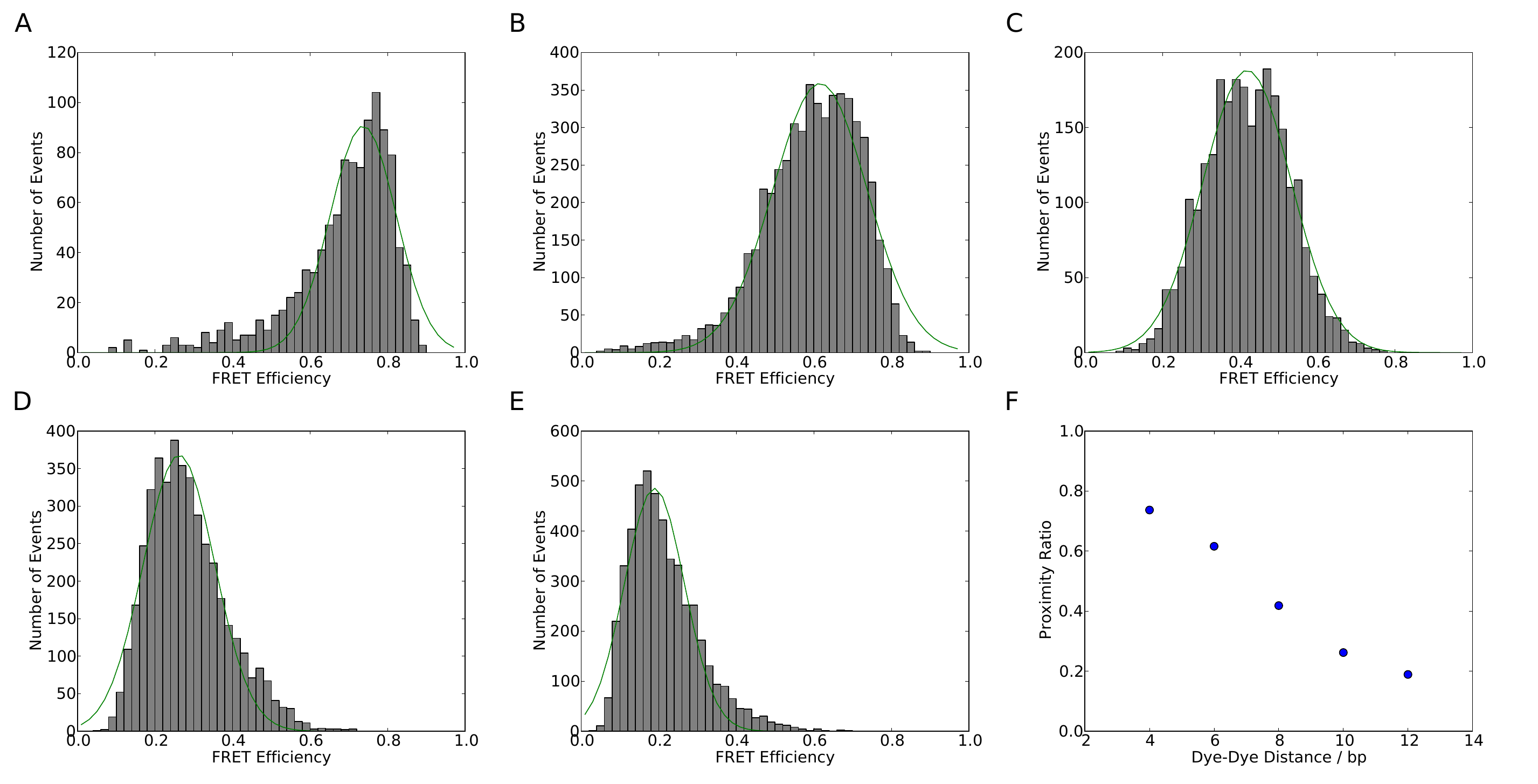}}
\caption{Analysis of FRET data from DNA duplexes using pyFRET. A - E: Fitted FRET histograms from DNA duplexes labelled with a dye-dye separation of 4, 6, 8, 10 and 12 base pairs respectively. F) Characteristic sigmoidal curve of FRET efficiency against dye-dye distance. \DUrole{label}{fig3FRET}}
\end{figure}\begin{figure}[]\noindent\makebox[\columnwidth][c]{\includegraphics[width=\columnwidth]{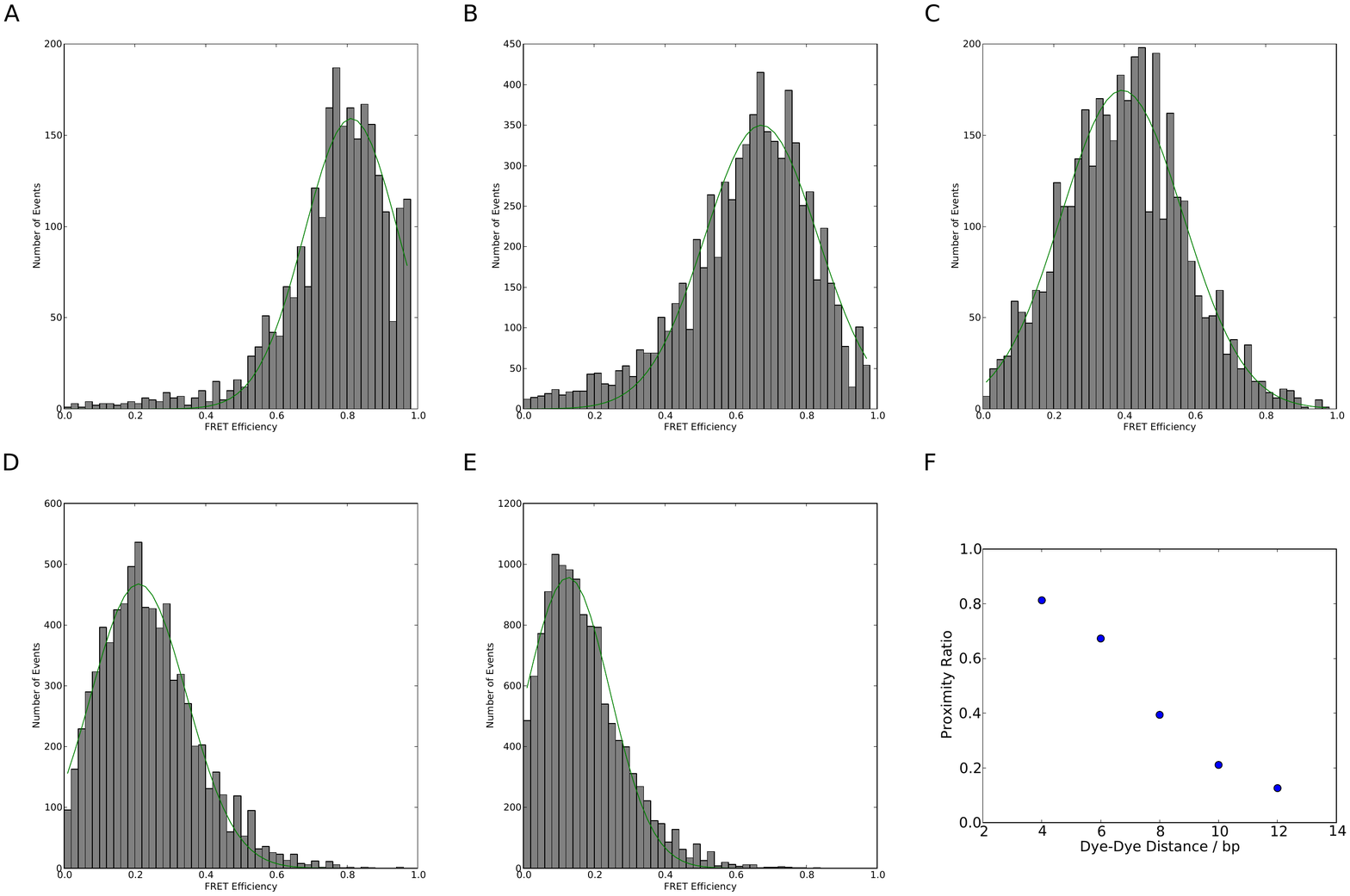}}
\caption{Analysis of ALEX data from DNA duplexes using pyALEX. A - E: Fitted FRET histograms from DNA duplexes labelled with a dye-dye separation of 4, 6, 8, 10 and 12 base pairs respectively. F) Characteristic sigmoidal curve of FRET efficiency against dye-dye distance. \DUrole{label}{fig3ALEX}}
\end{figure}

\section{Conclusion%
  \label{conclusion}%
}
pyFRET is available to download from PyPI under an open source three-clause BSD licence.
The source code is available from Bitbucket \cite{bitbucket}.
Documentation can also be found there, whilst a more extensive tutorial, including example scripts, can be found on our website at ReadTheDocs \cite{RTD}.

pyFRET currently provides basic tools for burst selection and denoising, based on simple thresholding and noise subtraction techniques.
We are aware that more sophisticated methodologies exist and are currently working to produce and open source burst selection algorithm based on photon arrival times \cite{nir06} as well as stochastic denoising algorithms \cite{kudryavtsev2012}.
We have also developed a novel analysis method based on Bayesian statistics \cite{murphy14}, for which source code is available (\url{https://bitbucket.org/rebecca_roisin/fret-inference}\}) and which will be folded into the pyFRET library.
We are also working to increase support for the wide variety of file formats that result from custom-built data collection hardware.

smFRET is a fast-developing and active research field and we want to support scientific progress through development of high-quality usable software.
We are keen to work with others to enable their use of and contribution to the pyFRET library.
We welcome requests for custom analysis requirements and are happy to support others who wish to contribute additional code to the pyFRET infrastucture.

\end{document}